\documentclass[a4paper,dvips,12pt]{article}
\input{axodraw.sty}
\usepackage{amsmath,amssymb,exscale}   
\usepackage{array,multicol}
\usepackage{afterpage,float,flafter}   
\usepackage{epsfig,rotating,pifont,fancybox}   
\usepackage{cite}
\makeatletter   
\@addtoreset{equation}{section}   
\makeatother   
   
\newcommand{\bea}{\begin{eqnarray}}   
\newcommand{\eea}{\end{eqnarray}}   
   
\newcommand{\NPB}[3]{\emph{ Nucl.~Phys.} \textbf{B#1} (#2) #3}   
\newcommand{\PLB}[3]{\emph{ Phys.~Lett.} \textbf{B#1} (#2) #3}   
\newcommand{\PRD}[3]{\emph{ Phys.~Rev.} \textbf{D#1} (#2) #3}   
\newcommand{\PRL}[3]{\emph{ Phys.~Rev.~Lett.} \textbf{#1} (#2) #3}

\newcommand{\PR}[3]{\emph{ Phys.~Rep.} \textbf{#1} (#2) #3}

\newcommand{\JHEP}[3]{\emph{ JHEP} \textbf{#1} (#2) #3}

\title{   
\vspace*{-0.8cm}   
\begin{flushright}   
\normalsize{      
IEM-FT-222/02\\
\texttt{hep-th/0204041}}\\ 
\end{flushright}    
\vspace{1cm}
\Large{\sc Radiative Scherk-Schwarz supersymmetry
breaking~\footnote{Work supported in part by CICYT, Spain, under
contract FPA2001-1806, and by EU under contracts HPRN-CT-2000-00152,
HPRN-CT-2000-00148 and HPRN-CT-2000-00149.}}
\vspace*{.5cm}
\author{\large
{\sc G.~v.~Gersdorff~$^a$, M.~Quir{\'o}s~$^a$ and A.~Riotto~$^b$}\\ \\
\emph{~$^a$Instituto de Estructura de la Materia (CSIC), Serrano 123}\\
\emph{E-28006-Madrid, Spain}\\ 
\emph{~$^b$INFN, sezione di Padova, Via Marzolo 8,
I-35131 Padua, Italy}}
}
\date{}   
\begin{document}
\maketitle
\thispagestyle{empty}
\vspace*{.5cm}

\begin{abstract}\noindent
We analyze the Scherk-Schwarz (SS) supersymmetry breaking in
brane-world five dimensional theories compactified on the orbifold
$S^1/\mathbb{Z}_2$. The SS breaking parameter is undetermined at the
tree-level (no-scale supergravity) and can be interpreted as the
Hosotani vacuum expectation value corresponding to the $U(1)_R$ group
in five dimensional $N=2$ (ungauged) supergravity.  We show that the
SS breaking parameter is fixed at the loop level to either $0$ or
$1/2$ depending on the matter content propagating in the bulk but in a
rather model-independent way. Supersymmetry breaking is therefore
fixed through a radiative Scherk-Schwarz mechanism.  We also show that
the two discrete values of the SS parameter, as well as the
supersymmetry breaking shift in the spectrum of the bulk fields, are
altered in the presence of a brane-localized supersymmetry breaking
arising from some hidden sector dynamics. The interplay between the SS
and the brane localized breaking is studied in detail.
\end{abstract}
\vspace{2.5cm}   
   
\begin{flushleft}   
April 2002 \\   
\end{flushleft}
\newpage
\section{\sc Introduction}
\label{introduction}

If supersymmetry plays an important role in constructing consistent
high-energy physics theories, a relevant issue is to explain how
supersymmetry is spontaneously broken in the low-energy world. Recent
ideas on extra-dimensions and the speculation that our visible
universe coincides with a four-dimensional brane living in the bulk of
the extra-dimensions -- the so-called brane-world scenarios -- have
given rise to new appealing possibilities regarding how to realize
supersymmetry breaking~\cite{anton}--\cite{gp}. The extra dimensional
framework is particularly interesting because it provides a new
geometrical perspective in understanding some of the problems of
conventional four-dimensional theories.

In a recent paper~\cite{geromariano} five dimensional (5D)
supersymmetric theories compactified on the orbifold
$S^1/\mathbb{Z}_2$ were considered and the Scherk-Schwarz (SS)
supersymmetry breaking~\cite{ss} was interpreted as the Hosotani
breaking~\cite{hosotani} of the local $SU(2)_R$ symmetry present in
off-shell $N=2$ supergravity.  In particular it was shown that
$SU(2)_R$ is gauged by auxiliary fields $\vec{A}_M$, $M=\mu,5$ whose
extra-dimensional components background $V_5^1+i\,V_5^2$ -- that make
part of the $F$-component of the radion multiplet -- provide different
bulk masses to fermions and bosons in $N=1$ supermultiplets and then
trigger Scherk-Schwarz supersymmetry
breaking~\cite{geromariano,radionF}. The vacuum expectation value
(VEV) of the field $V_5^1+i\,V_5^2$ is however not fixed at the tree
level, a reminiscent situation of no-scale supergravity in four
dimensions~\cite{no-scale}, which means that the scale at which
supersymmetry is broken is classically undetermined.

In this paper we want to go one step further and introduce one-loop
radiative corrections to dynamically determine the Scherk-Schwarz
supersymmetry breaking parameter.  Our goal is to show that the
Scherk-Schwarz supersymmetry breaking parameter $\omega$ can be led by
radiative corrections to two discrete values, either 0 or 1/2,
corresponding to unbroken or broken supersymmetry.  The actual value
of $\omega$ depends upon the matter content of the theory in the bulk.
In order to do the analysis it is convenient to deal with only
physical degrees of freedom, integrate out the auxiliary fields and
work with on-shell $N=2$ supergravity. This will be done in
section~\ref{bulk} where the on-shell version of the mechanism in
Ref.~\cite{geromariano} will be presented. The one-loop effective
potential will be computed in section~\ref{radiative} where the
Scherk-Schwarz breaking will be dynamically determined. Non-local and
spontaneous supersymmetry breaking can also be induced by
brane-localized dynamics giving rise to effective supersymmetry
breaking superpotentials~\cite{BFZ,GR}. The interplay between this
mechanism and that induced by the $V_5^1+i\,V_5^2$ VEV is studied in
section~\ref{brane} that gives rise to a variety of possibilities for
the supersymmetry breaking shift in the spectrum of the bulk fields
depending on the actual value of the supersymmetry breaking
superpotential. Finally we draw our conclusions in
section~\ref{conclusion}.

\section{\sc Bulk supersymmetry breaking}
\label{bulk}

In this section we want to introduce the Scherk-Schwarz mechanism for
supersymmetry breaking in the context of on-shell $D=5$ $N=2$
supergravity, the on-shell version of the mechanism presented in
Ref.~\cite{geromariano}.  $D=5$ $N=2$ supergravity compactified on the
$\mathbb{Z}_2$ orbifold has been recently analyzed in
Refs.~\cite{bergshoeff,bagger} and given an off-shell formulation in
Refs.~\cite{kugo,zucker}. The starting point is the minimal
supergravity multiplet in five dimensions containing the graviton
$g_{MN}$, the $SU(2)_R$ doublet gravitino $\psi^a_M$ and the
graviphoton $B_M$ as propagating fields and, as auxiliary fields, the
$SU(2)_R$ gauge fields $\vec{V}_M$, an antisymmetric tensor $v^{AB}$,
an $SU(2)_R$ triplet $\vec{t}$, a real scalar $C$ and an $SU(2)_R$
doublet spinor $\zeta^a$. The orbifold parity assignments are
displayed in Table~\ref{paritymin}.
\begin{table}[htb]
\begin{center}
\begin{tabular}{||c|c|c||}
\hline\hline
{\sc field}  &{\sc even}&{\sc odd}\\
\hline
$g_{MN}$ & $g_{\mu\nu}$, $g_{55}$&$g_{\mu 5}$\\
$\psi_M$& 
        $\psi_{\mu L}^1$, $\psi_{5L}^2$&$\psi_{\mu L}^2$, 
	$\psi_{5 L}^1$\\
$B_M$&
        $B_5$&$B_\mu$\\
$\vec V_M$&$V_\mu^3$, $V_5^{1,2}$&$V_\mu^{1,2}$, $V_5^3$\\
$v^{AB}$& $v^{\alpha 5}$&
	$v^{\alpha \beta}$\\
$\vec t$ &$t^{1,2}$&$t^3$\\
$C$&$C$&\\
$\zeta$&$\zeta_L^1$&$\zeta_L^2$\\
\hline\hline
\end{tabular}
\end{center}
\caption{Parity assignment  of the minimal supergravity multiplet in
five dimensions.}
\label{paritymin}
\end{table}
However the action based on the minimal multiplet is not physical, as
was observed in~\cite{zucker}. A simple way of realizing this fact is
that it contains $40_B+40_F$ degrees of freedom while the minimal
$D=5$ off-shell supergravity must contain at least $48_B+48_F$ degrees
of freedom~\cite{sohnius}. Another way of realizing that the minimal
multiplet by itself does not describe the physical on-shell action is
the variation of the action with respect to $C$ which leads to the
unphysical equation of motion $\det g_{MN}=0$. Therefore introduction
of an additional (compensator) $8_B+8_F$ multiplet is required.  This
compensator will serve to (partially) fix the $SU(2)_R$ local symmetry
in the on-shell theory.  Depending on the choice of the additional
supermultiplet off-shell supergravity looks different although the
physical on-shell versions are the same for all cases. Different
choices have been done in Refs.~\cite{zucker}: a nonlinear multiplet
(version I), a hypermultiplet (version II) and a tensor multiplet
(version III). Here we will adopt the latter formulation of off-shell
supergravity and introduce an additional tensor supermultiplet. In
this case it is not possible to fix the whole $SU(2)_R$ symmetry by
means of the compensator field $\vec{Y}$ but a residual $U(1)_R$
gauged by an auxiliary field remains. However this residual degree of
freedom can be fixed by means of the equations of motion of the
compensating multiplet.

The $D=5$ tensor multiplet contains an $SU(2)_R$ bosonic triplet
$\vec{Y}$, an antisymmetric three-form potential $B_{MNP}$, a real
scalar $N$ and an $SU(2)_R$ doublet spinor $\rho^a$.  The orbifold
parity assignments of the tensor multiplet are displayed in
Table~\ref{parityten}.
\begin{table}[htb]
\begin{center}
\begin{tabular}{||c|c|c||}
\hline\hline
{\sc field}  &{\sc even}&{\sc odd}\\
\hline
$\vec{Y}$ & $Y^1,\, Y^2$& $Y^3$\\
$B_{MNP}$ & 
        $B_{\mu\nu\rho}$ & $B_{\mu\nu 5}$ \\
$N$ &
        $N$ &  \\
$\rho$ & $\rho_L^1$ &$\rho_L^2$ \\
\hline\hline
\end{tabular}
\end{center}
\caption{Parity assignment  of the tensor multiplet.}
\label{parityten}
\end{table}
The $SU(2)_R$ gauge fixing is done by the compensator field $\vec{Y}$ as
\begin{equation}
\label{gfix}
\vec{Y}=e^u
\left(\begin{array}{c}
0\\
1\\
0
\end{array}
\right)
\end{equation}
where the scalar field $u$ has been introduced. This gauge leaves the $U(1)_R$
subgroup generated by $\sigma^2$ unbroken. Now, if we take as the Lagrangian
\begin{equation}
\label{lag}
\mathcal{L}_{grav}=\mathcal{L}_{minimal}+\mathcal{L}_{tensor}\, ,
\end{equation}
the terms proportional to $C$ in $\mathcal{L}_{grav}$ are $\sim
(1-e^u)C$. Variation with respect to $C$ now yields the equation of
motion $u=0$, which shows that the previously mentioned problem of
$\mathcal{L}_{minimal}$ is solved.  The VEV $u=0$ will be fixed
hereafter.

The auxiliary fields that are relevant for supersymmetry breaking are
$V_5^1$, $V_5^2$, $t^1$ and $t^2$, the components of the $F-$term of
the radion multiplet
\begin{equation}
\label{radion}
\left[h_{55}+i B_5, \psi_{5 L}^2,V_5^1+i V_5^2+4i(t^1+i t^2)
\right]\, .
\end{equation}
The relevant terms in $\mathcal{L}_{grav}$ containing 
$V_5^1$, $V_5^2$, $t^1$ and $t^2$ are
\begin{align}
\label{lagaux}
\mathcal{L}_{grav}=&-\frac{i}{2}\bar\psi_P \gamma^{PMN}\mathcal{D}_M\psi_N
-\frac{1}{12}\varepsilon^{MNPQR}V_M^2\partial_N B_{PQR}\nonumber\\
+&(V_5^1)^2-12 (t^1)^2-48(t^2)^2-12 Nt^2-N^2\, ,
\end{align}
where $\mathcal{D}_M$ is the covariant derivative with respect to local 
Lorentz and local $U(1)_R$ transformations
\begin{equation}
\label{covder}
\mathcal{D}_M=D_M+i\,\sigma^2 V_M^2
\end{equation}
and $D_M$ the covariant derivative with respect to local Lorentz 
transformations.

The variation of (\ref{lagaux}) with respect to auxiliary fields
provides the on-shell supergravity Lagrangian. In general, auxiliary
fields $X$ have couplings in $\mathcal{L}_{aux}$, whose variation
provides the vacuum expectation value (VEV) $X_0$, and couplings to
matter (propagating) spinor $\psi$ and scalar $A$ fields as
$\mathcal{L}_{int}=a_F X \bar\psi \psi+a_B X^2 |A|^2$.  If we are only
interested in mass terms we can replace $X\to X_0$ into
$\mathcal{L}_{int}$ since the difference when solving the field
equations from $\mathcal{L}_{aux}$ and
$\mathcal{L}_{aux}+\mathcal{L}_{int}$ are quartic terms in matter
fields.  In particular, the field equations for the auxiliary fields
$V_5^1$, $N$, $t^1$ and $t^2$ yield
\begin{equation}
\label{solucion1}
\langle V_5^1\rangle=\langle N\rangle=\langle t^1\rangle=\langle t^2
\rangle=0\, ,
\end{equation}
while the field equation for the three-form tensor field $B_{MNP}$ gives
\begin{equation}
\label{3forma}
\partial_{[M}V^2_{N]}=0\, .
\end{equation}
The obvious solution to the latter equation is 
\begin{equation}
\label{solucion2}
V_M^2=\partial_M C\, ,
\end{equation}
{\it i.e.\ }a pure gauge that is non-trivial in a compact space. Since
$C$ is odd, the simplest choice
\begin{equation}
C= \omega\,\frac{x^5}{R}
\end{equation}
leads to the background
\begin{equation}
V_\mu^2=0
\end{equation}
and 
\begin{equation}
V_5^2=\frac{\omega}{R}
\end{equation}
that breaks supersymmetry through the coupling (\ref{covder}) in
(\ref{lagaux}) as in Ref.~\cite{geromariano}.

Using the coupling of $V_5^2$ to the gravitino field in (\ref{lagaux})
one obtains the gravitino mass eigenvalues for the Kaluza-Klein modes
in the orbifold compactification $S^1/\mathbb{Z}_2$
\begin{equation}
\label{masa32}
m_{3/2}^{(n)}=\frac{n+\omega}{R}\, ,
\end{equation}
where $n$ is an integer number. Supersymmetry breaking is also
manifest in the spectrum of bulk vector multiplets and scalar
hypermultiplets.  The action for the super Yang-Mills fields and
hypermultiplets have been computed in Ref.~\cite{zucker} by embedding
the corresponding $D=5$ $N=2$ supermultiplets into linear
multiplets. The relevant couplings of the $SU(2)_R$ doublets, gauginos
$\lambda$ and hyperscalars $A$, with the $SU(2)_R$ gauge bosons
$\vec{V}_M$ is provided by the covariant derivative in the interaction
Lagrangian,
\begin{equation}
\label{lagmateria}
\mathcal{L}_{matter}=\frac{i}{2} \bar\lambda\gamma^M \widehat{\mathcal{D}}_M
\lambda+\left|\widehat{\mathcal{D}}_M A\right|^2\, ,
\end{equation}
where 
\begin{equation}
\label{covderhat}
\widehat{\mathcal{D}}_M=D_M+i\,\vec{\sigma}\vec{V}_M
\end{equation}
is the original covariant derivative with respect to the whole
$SU(2)_R$ gauge group.  After using the field equations for the
auxiliary fields $V_5^1=V_5^3=0$ we obtain that
$\widehat{\mathcal{D}}_M\to\mathcal{D}_M$ and the mass eigenvalues for
gauginos and hyperscalars can be computed as in
Ref.~\cite{geromariano} and yield
\begin{equation}
\label{masa120}
m_{1/2}^{(n)}=m_{0}^{(n)}=\frac{n+\omega}{R}\, .
\end{equation}
All the mass eigenvalues depend on the field $V_5^2$ in the same way.
Therefore the mass of zero modes for gauginos, gravitinos and
hyperscalars is $m^0=V_5^2$ which is then the scale of supersymmetry
breaking. This scale remains undetermined at the tree level, a
behaviour typical of Scherk-Schwarz breaking and reminiscent of
no-scale supergravity models~\cite{no-scale}. This flatness will be
lifted when loop corrections will be accounted, thus leading to the
possibility of breaking radiatively supersymmetry through the
Scherk-Schwarz mechanism. This is the subject of the next section.

\section{\sc Radiative determination of supersymmetry \\ breaking}
\label{radiative}

The one-loop effective potential can be most easily computed using 5D
functional techniques. In particular for hyperscalars it is given by
\begin{equation}
\label{efpot0}
V_0=\frac{2N_H}{2}\, \int\frac{d^5 p}{(2\pi)^5}\ln \left(p^2+p_5^2\right)\, ,
\end{equation}
where $N_H$ is the number of hypermultiplets, the factor of $2$ comes
from the degrees of freedom of a complex scalar and $\int dp_5/2\pi\to
\sum_{p^5}$ is understood with $p_5=m_0^{(n)}$.  For gauginos it is
\begin{equation}
\label{efpot12}
V_{1/2}=-\frac{2N_V}{2}\,  
\int\frac{d^5 p}{(2\pi)^5}\ln \left(p^2+p_5^2\right)\, ,
\end{equation}
where $N_V$ is the number of vector multiplets, the factor of $2$
comes from the degrees of freedom of a Majorana gaugino and
$p_5=m_{1/2}^{(n)}$.  Both in (\ref{efpot0}) and (\ref{efpot12}) we
have already taken into account a global factor of $1/2$ for the
orbifold $S^1/\mathbb{Z}_2$ compactification that removes half of the
degrees of freedom corresponding to the circle.

The contribution of the gravitino is easily obtained using the
functional techniques of Refs.~\cite{gravitinoeff}. In particular one
can see that in the 5D harmonic gauge $\gamma^N \psi_N^a=0$ the one
loop effective potential for the 5D gravitino can be easily worked
out. It gives,
\begin{equation}
\label{efpot32}
V_{3/2}=-\frac{4}{2}\,\int\frac{d^5 p}{(2\pi)^5}\ln \left(p^2+p_5^2\right)\, ,
\end{equation}
where a 5D gravitino has eight degrees of freedom on-shell, the factor
$4$ appears from the orbifold action and $p_5=m_{3/2}^{(n)}$.

Using now the mass eigenvalues for gravitinos (\ref{masa32}) and
gauginos and hyperscalars (\ref{masa120}), the one-loop effective
potential in the presence of the background $V_5^2=\omega/R$ for a
system of $N_H$ hyperscalars and $N_V$ vector multiplets propagating
in the bulk of the fifth dimension can be computed as~\cite{ADPQ}
\begin{equation}
\label{potencial}
V_{eff}(\omega)=\frac{3(2+N_V-N_H)}{64\pi^6 R^4}\, 
\left[{\rm Li}_5\left(e^{2 i \pi \omega}
\right)+{\rm h.c.}\right]\, ,
\end{equation}
where ${\rm Li}_n(x)$ are the polylogarithm functions
$${\rm Li}_n(x)=\sum_{k=1}^{\infty}\frac{x^k}{k^n}.$$

Using the power expansion of ${\rm Li}_5(e^{2i \pi\omega})+
{\rm h.c.}$ around $\omega=0$,
\begin{equation}
\label{expan}
{\rm Li}_5\left(e^{2 i \pi \omega}\right)+{\rm h.c.}
=2 \, \zeta(5)-4\pi^2 \zeta(3)\,\omega^2+
\cdots\, ,
\end{equation}
we can see that the potential (\ref{potencial}) has a maximum
(minimum) at
\begin{equation}
\omega=0
\end{equation}
for $N_H<2+N_V$ ($N_H>2+N_V$). This means that for $N_H>2+N_V$ the
global minimum of the potential is at $\omega=0$ and there is no
induced supersymmetry breaking. On the other hand, for $N_H<2+N_V$ the
global minimum is at
\begin{equation}
\omega=1/2\, ,
\end{equation}
supersymmetry is broken radiatively and the KK spectrum of bulk
gauginos and hyperscalars is shifted from $n/R$ to $(n+1/2)/R$ . We
have dubbed this phenomenon radiative Scherk-Schwarz supersymmetry
breaking.  Notice that our findings are quite model-independent since
they depend only upon the total number of vector multiplets and
hypermultiplets in the bulk.

Some comments are now in order. The first one refers to radius
stabilization.  As one can see from (\ref{potencial}) the one-loop
potential does not stabilize the value of $R$, related to the radion
field. In fact the behaviour is runaway: in particular for
$N_H<2+N_V$, we obtain at the global minimum of the effective
potential the asymptotic behaviour $R\to 0 $. However we expect the
radion to be stabilized by some other mechanism that does not affect
the present results. Massive fields in the bulk seem to be required
for radion stabilization~\cite{radion}.

The second comment concerns the appearence of a non-vanishing vacuum
energy at the minimum of the effective potential of order $1/R^4$.
This fact is a generic feature of many models with flat potential, or
zero cosmological constant, at the tree-level. We expect on general
grounds that the non-vanishing energy will disturb our original
assumption of flat space-time. We are not computing the back reaction
on the space-time metrics but expect its modification to be suppressed
as $1/(M_5 R)^4$, where $M_5$ is the Planck scale of the higher
dimensional theory.

The last comment refers to the quality of the supersymmetry breaking
solution $\omega=1/2$ we have found. Since all the supersymmetric
spectrum depends on the actual value of $\omega$~\cite{ADPQ} the value
$\omega=1/2$ is very constraining for electroweak symmetry breaking in
the different considered models. In the model presented in
Ref.~\cite{ADPQ}, where all matter fields were localized on the branes
while Higgs and vector multiplets live in the bulk, the case
$\omega=1/2$ is phenomenologically problematic because the theory does
not allow for an $H_1H_2$ Higgs mixing and therefore the VEV of $H_1$
is zero. The same happens for the alternative model~\cite{DQ} where
gauge multiplets and $SU(2)_L$ singlets live and the bulk and
$SU(2)_L$ doublets are localized on the brane. This problem is however
avoided in models with a single Higgs hypermultiplet, as those based
on $S^1/\mathbb{Z}_2\times \mathbb{Z}'_2$~\cite{barbieri}, although
these models develop a Fayet-Iliopoulos quadratic
divergence~\cite{FI}.

We have checked as well that in models with $N_H<2+N_V$ the stability
of the minimum at $\omega=1/2$ is not lifted by two-loop
corrections. Then it would be desirable to introduce an extra
mechanism that would lead to an effective potential with a minimum at
$\omega\neq 0,1/2$. Since the orbifold $S^1/\mathbb{Z}_2$ has branes
at the fixed points, a hidden sector at one of the fixed points (say
at $x^5=0$) can break spontaneously supersymmetry by some
brane-localized dynamics, as {\it e.g.}  gaugino condensation.  In
this case there will be an interplay between the mechanism that we
have discussed up to now and the brane dynamics. This will be the
object of study in the next section.

\section{\sc Brane-assisted supersymmetry breaking}
\label{brane}

The couplings of the minimal and tensor multiplet to the branes have
been analyzed in Ref.~\cite{zucker}. Using the gauge (\ref{gfix}) they
lead to the total Lagrangian,
\begin{equation}
\label{total}
\mathcal{L}=\mathcal{L}_{grav}+W\,\delta(x^5)\,\mathcal{L}_{brane}\, ,
\end{equation}
where $W$ is a constant superpotential left out when some brane fields in the
hidden sector are integrated out and
\begin{equation}
\label{lagbrana}
\mathcal{L}_{brane}=-2\,N-2\,V_5^1-12\, t^2
+\frac{1}{2}\bar\psi_M \sigma^2\gamma^{MN}\psi_N+\cdots\, ,
\end{equation}
where we have fixed the VEV of the auxiliary fields that do not play any
role in the mass spectrum of $SU(2)_R$ doublets. 

We are assuming the hidden sector in the brane located at $x^5=0$ and
therefore the observable brane (where any localized matter fields
should be located) at $x^5=\pi R$~\footnote{Of course the choice of
hidden and observable branes is absolutely arbitrary and can be
exchanged by a coordinate redefinition.}. Notice that the gauge fixing
condition (\ref{gfix}) breaks $SU(2)_R$ in the bulk to the $U(1)_R$
generated by $\sigma^2$, and to nothing in the branes since only the
auxiliary gauge field $A_\mu^3$ is even (see
Table~\ref{paritymin}). The bulk $U(1)_R$ will drop out by elimination
of the auxiliary field $V_5^2$ from the $B_{MNP}$ field equation
(\ref{solucion2}).

However the presence of the brane Lagrangian (\ref{lagbrana}) will
provide a brane mass contribution to the 5D gravitino and will alter
the field equations of $V_5^1$, $t^1$ and $t^2$ with respect to the
bulk solution (\ref{solucion1}). This in turn will modify the spectrum
of mass eigenvalues of gauginos and hyperscalars as we will see in
this section.

The field equation for the auxiliary fields $V_5^1$, $N$, $t^1$ and $t^2$ from 
the Lagrangian (\ref{total}) in the presence of the brane term 
(\ref{lagbrana}) lead to
\begin{align}
\label{solucionb}
\langle t^1\rangle &=\langle t^2\rangle=0\, ,\nonumber\\
\langle N\rangle &=-W\delta(x^5)\, ,\nonumber\\
\langle V_5^1\rangle &=W\delta(x^5)\, .
\end{align}
After elimination of the auxiliary fields $V_5^1$, $N$, $t^1$ and $t^2$ by
their field equations (\ref{solucionb}) there appear terms of the form
$\delta^2(x^5)$. We have checked
that these singular terms  
cancel leading to a zero cosmological constant at the
tree-level. This latter 
property is expected to be spoiled by radiative corrections
as we have discussed in the previous section.

The presence of the gravitino brane mass term in (\ref{lagbrana}) as
well as the non-trivial solution for the field $V_5^1$ in
(\ref{solucionb}) will alter the mass spectrum for gravitinos,
gauginos and hyperscalars, whose mass eigenvalues will departure from
the calculated values in (\ref{masa32}) and (\ref{masa120}), and need
to be recalculated.  The Kaluza-Klein spectrum of bulk vector
multiplets, hyperscalars and gravitinos in the presence of the
auxiliary fields $V_5^2= \omega/R$ and $V_5^1=W\delta(x^5)$ can be
computed in different ways. One can either directly solve for the
corresponding equations of motion in $D=5$ or diagonalize the
corresponding infinite mass matrices. In this paper, we adopt another
strategy, {\it i.e.} the Dyson resummation.

\subsection{\sc Gauginos}

We can expand gauginos in modes according to the parity action
$$\lambda(x^\mu,x^5)=i\gamma^5 \cdot \sigma^3\lambda(x^\mu,-x^5)\, ,$$
{\it i.e.} $\lambda^1_L$ even and $\lambda^2_L$ odd, as
\begin{equation}
\label{gaugidec}
\left(
\begin{array}{c}
\lambda^1_{L}\\ 
\lambda^2_{L}
\end{array}
\right)
=\frac{1}{\sqrt{\pi R}}\sum_{n=-\infty}^{\infty}\lambda_{L}^{(n)}
\left(
\begin{array}{l}
\cos n x^5/R \\
\sin n x^5/R
\end{array}
\right)\, .
\end{equation}
The mass term for gauginos from (\ref{total}), taking into account the
classical VEV of $V_5^1$ in (\ref{solucionb}), can be written as
\begin{equation}
\label{matriz12}
\mathcal{M}_{1/2}=i\gamma^5 \widehat{\mathcal{D}}_5=
i\gamma^5\left(\partial_5 +i\, \sigma^1 \, V_5^1 +i\,\sigma^2 \,
V_5^2\right)\, .
\end{equation}
The mode expansion (\ref{gaugidec}) leads to
\begin{equation}
\label{masagauge}
\mathcal{L}_{m_{1/2}}=\frac{1}{2}\sum_n\frac{n+\omega}{R}\bar\lambda^{(n)} 
\lambda^{(n)}+\frac{W}{\pi R}\sum_{n,m}\bar\lambda^{(n)}\gamma^5 
\lambda^{(m)}
\end{equation}
where $\lambda^{(n)}$ is defined as a four-component Majorana
spinor. The appearance of $\gamma^5$ in the second term of
(\ref{masagauge}) comes from the fact that the (brane) contribution of
$V_5^1$ in the covariant derivative $\widehat{\mathcal{D}}_5$ to the
gaugino mass is imaginary with respect to that of $V_5^2$.

The (infinite) mass matrix for modes $\lambda^{(n)}$ arising from
(\ref{masagauge}) is a very complicated one. A simple way of
diagonalizing it is by considering the first term of (\ref{masagauge})
in the unperturbed propagator
\begin{center}%
\SetScale{0.6}
\begin{picture}(70,30)(0,28)
\Line(20,50)(90,50)
\end{picture}\hspace{-.5cm}
$\equiv{\displaystyle \sum_{m,n}}\ n$
\SetScale{0.6}
\begin{picture}(70,30)(0,28)\hspace{-.5cm}
\Line(20,50)(90,50)
\end{picture}\hspace{-.8cm}$m={\displaystyle \sum_{m,n}}\ {\displaystyle 
\frac{i}{\not\! p-{\displaystyle \frac{n+\omega}{R}}}
\ \delta_{nm}}=i\,\pi R \cot\left(\not\! p\pi R-\pi\omega\right)$
\end{center}
where $\not\! p=\gamma^\mu p_\mu$, and the second term as a perturbation 
\begin{center}%
$n$
\SetScale{0.6}
\begin{picture}(70,30)(0,28)\hspace{-.5cm}
\Line(20,50)(80,50)
\Vertex(80,50){5}
\Line(85,50)(145,50)
\end{picture}$\quad m={\displaystyle-i\, \frac{W}{\pi R}}\gamma^5$
\end{center}
that can be introduced in the propagator by a Dyson resummation.

We will then compute the fermion propagator in the bulk but with end
points on the brane $x^5=0$,
\begin{center}
$\left<\lambda(p^\mu,x^5=0)\bar\lambda(p^\mu,x^5=0) \right>
={\displaystyle \sum_{n,m}}
\left<\lambda^{(n)}(p^\mu)\bar\lambda^{(m)}(p^\mu)\right>\equiv$

\SetScale{0.6}
\begin{picture}(50,30)(0,28)\hspace{-.5cm}
\Line(20,50)(80,50)
\end{picture}
\hspace{-.65cm}
+
\SetScale{0.6}
\begin{picture}(50,30)(0,28)\hspace{-.5cm}
\Line(20,50)(80,50)
\Vertex(80,50){5}
\Line(85,50)(145,50)
\end{picture}\hspace{1.5cm}
\hspace{-.65cm}
+
\SetScale{0.6}
\begin{picture}(50,30)(0,28)\hspace{-.5cm}
\Line(20,50)(80,50)
\Vertex(80,50){5}
\Line(85,50)(145,50)
\Vertex(145,50){5}
\Line(150,50)(210,50)
\end{picture}$\hspace{2.5cm}+\quad\cdots\ \equiv$
\end{center}
\begin{equation}
\label{resumgaug}
\pi R\, \cot\left(\not\! p\pi R-\pi\omega\right)\,
\frac{i}{\tan\left(\not\! p\pi R-\pi\omega\right)-W\gamma^5}\,
\tan\left(\not\! p \pi R-\pi\omega\right)\, .
\end{equation}
The mass eigenvalues are then computed as the poles of
the propagator in the particle rest frame, $p^\mu=(m,\vec{0})$, 
{\it i.e.}
\begin{equation}
\label{polegaug}
\det\left[\tan\left(m\gamma^0 \pi R-\pi\omega\right)-W\gamma^5\right]\, ,
\end{equation}
leading to 
\begin{align}
\label{masapolegaug}
m_{1/2}^{(n)}&=\frac{n+\Delta_{1/2}}{R}\, ,\nonumber\\
\Delta_{1/2}&=\frac{1}{\pi}\arctan
\sqrt{\frac{W^2+\tan^2\pi\omega}{1+W^2\tan^2\pi\omega}}\, .
\end{align}
For $W=0$ this reduces to the usual Scherk-Schwarz shift
$\Delta_{1/2}=\omega$ while for $\omega=0$ one finds
$\Delta_{1/2}=1/\pi\arctan W$.

\subsection{\sc Gravitinos}

Gravitinos can be expanded in modes, according with the parities of
Table~\ref{paritymin}, as
\begin{equation}
\label{gravdec}
\left(
\begin{array}{c}
\psi^1_{\mu L}\\ 
\psi^2_{\mu L}
\end{array}
\right)
=\frac{1}{\sqrt{\pi R}}\sum_{n=-\infty}^{\infty}\psi_{\mu L}^{(n)}
\left(
\begin{array}{l}
\cos n x^5/R \\
\sin n x^5/R
\end{array}
\right)\, .
\end{equation}
The mass terms for gravitinos from the Lagrangian (\ref{total}) including
brane terms in $\mathcal{L}_{brane}$ can be written as,
\begin{equation}
\label{masagrav}
\mathcal{L}_{m_{3/2}}=\frac{1}{2}\sum_n\frac{n+\omega}{R}\bar\psi^{(n)}_{\mu} 
\gamma^{\mu\nu}
\psi^{(n)}_{\nu}+\frac{W}{\pi R}\sum_{n,m}\bar\psi^{(n)}_{\mu} 
\gamma^{\mu\nu}
\psi^{(m)}_{\nu}\, ,
\end{equation}
where $\psi^{(n)}_{\mu}$ is defined as a four-component Majorana
spinor.

Diagonalization of the gravitino mass matrix can be done using
techniques similar to those used in the previous subsection for the
gauginos.  The main difference being that there is not any $\gamma^5$
factor in the second term of (\ref{masagrav}) and hence no $\gamma^5$
factor in the mass insertion between modes $\psi^{(n)}_{\mu}$ and
$\psi^{(m)}_{\nu}$. The Dyson resummation of the perturbation would
lead to the resummed propagator with poles (in the particle rest
frame) defined as solution of the equation,
\begin{equation}
\label{polegrav}
\det\left[\tan\left(m\gamma^0 \pi R-\pi\omega\right)-W\right]\, ,
\end{equation}
leading to 
\begin{align}
\label{masapolegrav}
m_{3/2}^{(n)}&=\frac{n+\Delta_{3/2}}{R}\, ,\nonumber\\
\Delta_{3/2}&=\omega+\frac{1}{\pi}\arctan W\, .
\end{align}

\subsection{\sc Hyperscalars}

The action of the $\mathbb{Z}_2$ parity on hyperscalars is defined as
$$A(x^\mu,x^5)=\sigma^3 A(x^\mu,-x^5)\, ,$$ {\it i.e.} $A^1$ is even
and $A^2$ odd. We can expand hyperscalars in modes as
\begin{equation}
\label{hyperdec}
\left(
\begin{array}{c}
A^1\\ 
A^2
\end{array}
\right)
=\frac{1}{\sqrt{\pi R}}\sum_{n=-\infty}^{\infty}A^{(n)}
\left(
\begin{array}{l}
\cos n x^5/R \\
\sin n x^5/R
\end{array}
\right)\, .
\end{equation}
The mass term for hyperscalars from (\ref{total}), taking into account
the classical VEV of $V_5^1$ in (\ref{solucionb}), can be written as
\begin{equation}
\label{matriz0}
\mathcal{M}_0^2=\widehat{\mathcal{D}}_5\widehat{\mathcal{D}}^5\, .
\end{equation}
The mode expansion (\ref{hyperdec}) leads to the mass Lagrangian,
\begin{equation}
\label{masahyper}
\mathcal{L}_0=-\sum_n \left(\frac{n+\omega}{R}\right)^2 \bar A^{(n)}A^{(n)}
-\sum_{m,n}\bar A^{(m)}\left[\delta(0)\ 
\frac{W^2}{\pi R}+i\,\frac{W}{\pi}\ \frac{m-n}{R^2}\right] A^{(n)}\, .
\end{equation}
Again, as happened for gauginos and gravitinos, the infinite mass
matrix for the modes $A^{(n)}$ arising from (\ref{masahyper}) is very
complicated and the simplest method for diagonalizing it is by
considering the first term of (\ref{masahyper}) as part of the
propagator
\begin{center}%
\SetScale{0.6}
\begin{picture}(70,30)(0,28)
\DashLine(20,50)(90,50){5}
\end{picture}\hspace{-.5cm}
$\equiv{\displaystyle \sum_{m,n}}\ n$
\SetScale{0.6}
\begin{picture}(70,30)(0,28)\hspace{-.5cm}
\DashLine(20,50)(90,50){5}
\end{picture}\hspace{-.8cm}$m={\displaystyle \sum_{m,n}}\ {\displaystyle 
\frac{i}{p^2-{\displaystyle \left(\frac{n+\omega}{R}\right)^2}}
\ \delta_{nm}}=$
\begin{equation}
D^\omega(p)=i\,\frac{\pi R}{2\, p}\left[ 
\cot\left(p\pi R-\pi\omega\right)+\cot\left(p\pi R+\pi\omega\right)
\right]
\end{equation}
\end{center}
and the second term as the perturbations 
\begin{center}%
$n$
\SetScale{0.6}
\begin{picture}(70,30)(0,28)\hspace{-.5cm}
\DashLine(20,50)(60,50){5}
\Vertex(60,50){4}
\DashLine(63,50)(103,50){5}
\end{picture}$\hspace{-.5cm} m={\displaystyle-i\,\delta(0) \frac{W^2}{\pi R}}$
\end{center}
\begin{center}%
$n$
\SetScale{.6}
\begin{picture}(70,30)(0,28)\hspace{-.5cm}
\DashLine(20,50)(105,50){5}
\Text(38,30)[c]{$\triangleright$}
\end{picture}$\hspace{-.5cm}
 m={\displaystyle-\, \frac{W}{\pi R}\frac{m+\omega}{R}}$
\end{center}
\begin{center}%
$n$
\SetScale{0.6}
\begin{picture}(70,30)(0,28)\hspace{-.5cm}
\DashLine(105,50)(20,50){5}
\Text(37,30)[c]{$\triangleleft$}
\end{picture}$\hspace{-.5cm} 
m={\displaystyle\, +\frac{W}{\pi R}\frac{n+\omega}{R}}$
\end{center}
that can be included in the propagator by a Dyson resummation.
Notice the appearance of the interaction proportional to
$$\delta(0)={\displaystyle \frac{1}{\pi R}\sum_n 1}\, ,$$ a common
feature in $S^1/\mathbb{Z}_2$ compactifications~\cite{mirpes}.  This
factor is necessary for the consistency of the theory, as we will see
next.

We will then compute the propagator in the bulk but with end points on
the brane $x^5=0$,
\begin{center}
$\left<\bar A(p^\mu,x^5=0)A(p^\mu,x^5=0) \right>
={\displaystyle \sum_{n,m}}
\left<\bar A^{(n)}(p^\mu)A^{(m)}(p^\mu)\right>\equiv$
%

\SetScale{.6}
\begin{picture}(50,30)(0,28)\hspace{-.5cm}
\DashLine(20,50)(80,50){5}
\end{picture}
\hspace{-.65cm}
+
\SetScale{0.6}
\begin{picture}(50,30)(0,28)\hspace{-.5cm}
\DashLine(20,50)(60,50){5}
\Vertex(60,50){4}
\DashLine(63,50)(105,50){5}
\end{picture}\hspace{1.5cm}
\hspace{-1.5cm}
+
\SetScale{0.6}
\begin{picture}(50,30)(0,28)\hspace{-.5cm}
\DashLine(20,50)(140,50){5}
\Text(32,30)[c]{$\triangleright$}
\Text(63,30)[c]{$\triangleright$}
\end{picture}
\hspace{.8cm}
+
\SetScale{0.6}
\begin{picture}(50,30)(0,28)\hspace{-.5cm}
\DashLine(20,50)(140,50){5}
\Text(32,30)[c]{$\triangleright$}
\Text(63,30)[c]{$\triangleleft$}
\end{picture}
$\hspace{1cm}+\quad\cdots\ \equiv$
\end{center}
\begin{equation}
\label{resumhyper}
\frac{D^\omega(p)}{1-W^2\cot(pR\pi-\pi \omega)\cot(pR\pi+\pi \omega)}\, .
\end{equation}

The mass eigenvalues are then computed as the poles of the propagator
in the particle rest frame, $p^\mu=(m,\vec{0})$, {\it i.e.}  the
solutions of the equation
\begin{equation}
\label{polehyper}
\tan(mR\pi-\pi \omega)\tan(mR\pi+\pi \omega)-W^2=0\, .
\end{equation}
Equation (\ref{polehyper}) can be readily solved as
\begin{align}
\label{masapolehyper}
m_{0}^{(n)}&=\frac{n+\Delta_{0}}{R}\, ,\nonumber\\
\Delta_{0}&=\Delta_{1/2}\, .
\end{align}
where $\Delta_{1/2}$ was defined in (\ref{masapolegaug}). The fact
that mass eigenvalues for gauginos and hyperscalars are the same is a
consequence of the fact that both matter fields behave identically
with respect to the initial $SU(2)_R$, {\it i.e.} with the covariant
derivative $\widehat{\mathcal{D}}_M$ while the gravitino is acted by
$SU(2)_R$ by means of the covariant derivative $\mathcal{D}_M$ as a
consequence of the gauge fixing in the tensor multiplet. Notice that
the presence of the factor $\delta(0)$ was required to get this
identity.

\subsection{\sc The one-loop effective potential}

The previous calculation of the mass shifts for KK modes,
$\Delta_{3/2,1/2,0}(\omega,W)$ shows that in the two extreme cases so
far in the literature $W=0$, i.e. no brane effects, and $\omega=0$,
i.e.  no consideration of the Scherk-Scharz breaking~\footnote{Observe
nevertheless that considering the Scherk-Scharz parameter vanishing is
not optional since it corresponds to a flat direction of the
corresponding tree-level potential.}, the KK mass spectrum for all
bulk fields is the same. However, unlike the case of $W$, that is
supposed to arise from some brane dynamics in the hidden sector and so
it is a free parameter for our analysis, the parameter $\omega$ should
be found from one-loop corrections. The output will in general lead to
distinct mass spectra that should be obtained upon minimization of the
effective potential.

\begin{figure}[htb]
\vspace{.8cm}
\centering
\epsfig{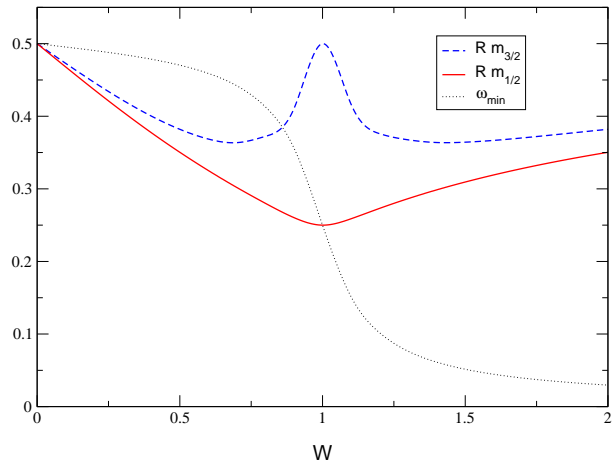}\vspace{.2cm}
\caption{\it Plot of the minimum of the effective potential $\omega_{min}$ 
(dotted line) and the corresponding lightest KK mode mass for the gravitino
(dashed line) and gaugino/hyperscalar (solid line) in units of $1/R$ for
the case of only Standard Model gauge bosons propagating in the bulk:
$N_V=12$, $N_H=0$.}
\label{fignothing}
\end{figure}

As we have seen above, the presence of the brane induced supersymmetry
breaking (\ref{lagbrana}) modified the mass eigenvalues for gravitinos
(\ref{masapolegrav}), gauginos (\ref{masapolegaug}) and hyperscalars
(\ref{masapolehyper}).  These in turn modify the one-loop effective
potential computed in section~\ref{radiative} that results now in
\begin{align}
\label{potencialbr}
V_{eff}(\omega)&=
\frac{3}{32\pi^6 R^4}\, \left[{\rm Li}_5
\left(e^{i 2 \pi \Delta_{3/2}(\omega,W)}
\right)+{\rm h.c.}\right]\nonumber\\
+&\frac{3(N_V-N_H)}{64\pi^6 R^4}\, \left[{\rm Li}_5\left(e^{i 2 \pi 
\Delta_{1/2}(\omega,W)}
\right)+{\rm h.c.}\right]\, .
\end{align}
We expect that the brane effects modify continuously the location of the
minimum of the effective potential away from its minima for $W=0$ at 
$\omega=0$ or $\omega=1/2$. We have studied numerically two extreme cases.

In Fig.~\ref{fignothing} we have studied the case where only the
gravitational and gauge sector are living in the bulk, while matter
and Higgs fields are localized in the observable brane and thus do not
participate in the one-loop effective potential. Of course the
supersymmetric partners of localized matter will receive radiative
masses from the bulk supersymmetry breaking~\cite{ADPQ}.  In
particular we have considered the supersymmetric Standard Model gauge
sector in the bulk with $N_V=12$ and $N_H=0$. The result is shown in
the plot where different quantities are shown vs. the parameter
$W$. The dotted line is the value of the Scherk-Schwarz parameter at
the minimum $\omega_{min}$. We can see that it is equal to $1/2$ for
$W=0$ and goes smoothly to zero in the limit $W\to\infty$. Dashed and
solid lines correspond to the values of the gravitino and
gaugino/hyperscalar lightest KK modes.
\begin{figure}[htb]
\vspace{.8cm}
\centering
\epsfig{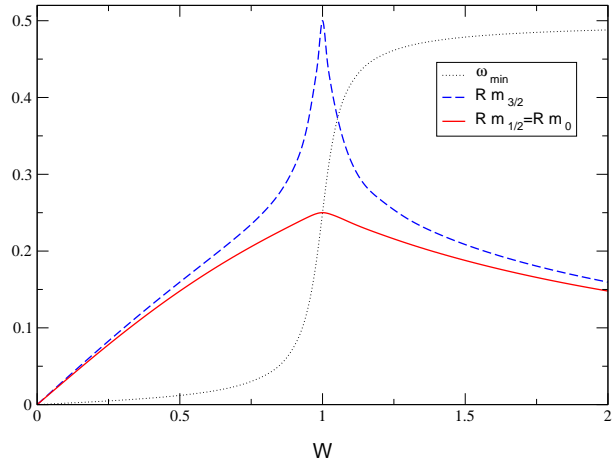}\vspace{.2cm}
\caption{\it The same as in Fig.~\ref{fignothing} for the case of all
Standard Model fields propagating in the bulk: $N_V=12$, $N_H=49$.}
\label{figall}
\end{figure}
Fig.~\ref{figall} corresponds to the other extreme case where all
gauge sector and matter (including two Higgs hypermultiplets)
propagate in the bulk. For the case of the supersymmetric Standard
Model $N_V=12$ and $N_H=3\times 15+4=49$. We can see that the minimum
starts at $\omega_{min}=0$ at $W=0$ and goes to $\omega_{min}=1/2$ in
the limit $W\to\infty$. In this limit the masses of lightest KK modes
can be much smaller than $1/R$.

\section{\sc Conclusions}
\label{conclusion}

In this paper we have considered a supersymmetric five-dimensional
brane-world scenario where the fifth dimension is compactified on
$S^1/\mathbb{Z}_2$. In this set-up, the Scherk-Schwarz supersymmetry
breaking can be interpreted as the Hosotani breaking of the $SU(2)_R$
local symmetry present in off-shell $N=2$ supergravity.  We have shown
that the Scherk-Schwarz supersymmetry breaking parameter is
undertemined at the tree-level, but can acquire two discrete values
after loop-corrections from supersymmetric bulk fields are
introduced. In this way, supersymmetry may get broken radiatively
through the Scherk-Schwarz mechanism. We have also computed the
contribution to the soft supersymmetry breaking masses of bulk fields
in the case in which a source of supersymmetry breaking appears
localized on the hidden brane. In such a case, the Scherk-Schwarz
supersymmetry breaking parameter is lifted away from the discrete
values $0$ and $1/2$ and the spectrum of the KK modes of the bulk
fields assumes a variety of possibilities depending upon the strength
of supersymmetry breaking on the hidden brane.


\section*{Acknowledgments} 
AR thanks the High-Energy Theory group of CSIC where part of this work
was done for the kind hospitality. MQ thanks the High-Energy Theory
group of University of Padova where this work was completed for the
kind hospitality.  One of us (AR) thanks F.~Feruglio for useful
discussions. The work of GG was supported by the DAAD.

\end{document}